\documentclass[showpacs,twocolumn,aps,prl]{revtex4}
\usepackage{amsmath}
\usepackage{amsfonts}
\usepackage{amssymb}
\usepackage{amsthm}
\usepackage{graphicx}

\begin{document}

\title{Critical current under an optimal time-dependent
polarization direction for Stoner particles in spin-transfer
torque induced fast magnetization reversal}
\author{Z. Z. Sun}
\affiliation{Physics Department, The Hong Kong University of
Science and Technology, Clear Water Bay, Hong Kong SAR, China}
\author{X. R. Wang}
\affiliation{Physics Department, The Hong Kong University of
Science and Technology, Clear Water Bay, Hong Kong SAR, China}
\date{\today}

\begin{abstract}
Fast magnetization reversal of uniaxial Stoner particles by
spin-transfer torque due to the spin-polarized electric current
is investigated. It is found that a current with a properly
designed time-dependent polarization direction can dramatically
reduce the critical current density required to reverse a
magnetization. Under the condition that the magnitude and
the polarization degree of the current do not vary with time,
the shape of the optimal time-dependent polarization direction
is obtained such that the magnetization reversal is the fastest.
\end{abstract}
\maketitle
The advent of miniaturization and fabrication of magnetic
particles at nano-meter scale\cite{Qikun} (called Stoner
particles) makes the Stoner-Wohlfarth (SW) problem\cite{book}
very relevant to nano-sciences and nano-technologies.
For a Stoner particle, strong exchange interactions keep the
magnetic moments of atoms rigid such that the constituent spins
rotate in unison. The magnetization dynamics can be manipulated
by a laser light\cite{Bigot}, or a magnetic field\cite
{field,xrw,xrw1}, or a spin-polarized electric current\cite
{Slon,theory,three,exp,sun,bauer,zhang,lee,switch}.
Among them, magnetization manipulation\cite{book} of Stoner
particles by electric current of spin-polarized electrons is of
great current interests in
spintronics because of its locality and low power consumption.
The idea of spin-transfer torque (STT) generated by a
spin-polarized electric current was independently suggested\cite
{Slon} by Slonczewski and Berger in 1996, and was verified by
several experiments\cite{exp}. Important issues in its
applications are to lower the critical current required to reverse
a magnetization\cite{sun} and to design a current pulse such that
the magnetization can be switched from one state to another fast.
Many reversal schemes\cite{xrw1,sun,bauer,zhang,lee,switch} have
been proposed and examined. Ideas include thermal-assistance\cite
{zhang,bauer} and sample designs\cite{sun,lee,switch}.
However, the question of how much the critical current can
further be lowered is still unknown. In this report, the optimal
time-dependence of the current polarization direction for the
fast magnetization reversal is derived when the magnitude
and the polarization degree of the current is fixed. It is
shown that the critical current density can be dramatically
reduced if the optimal time-dependent polarization is employed.
\begin{figure}[htbp]
 \begin{center}
\includegraphics[width=7.cm, height=4.cm]{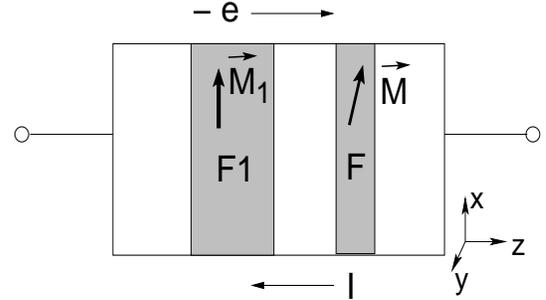}
 \end{center}
\caption{\label{fig1} Schematic illustration of the STT structure.
Note that the direction of the electrical current is opposite to
that of electron flow. }
\end{figure}

The prototype of STT systems is a magnetic multilayered structure
of nano-meter scale as illustrated in Fig. 1. It consists of two
ferromagnets which are sandwiched among three nonmagnetic metallic
layers. Electrons travel along the $+\hat{z}$ direction (from the
left to the right) in the sample (the direction of the current is
opposite as shown in Fig.1). The first ferromagnet F1 is usually
very thick so that the current does not affect the magnetization
$\vec{M}_1$ of F1. Electrons are polarized along $\vec{M}_1$ after
they pass through F1, and remain their polarizations before entering
the second ferromagnet F when the thickness of the spacer layer
between F1 and F is much smaller than the spin diffusion length.
The polarized electrons transfer their spin angular momenta to F,
resulting in so-called STT\cite{Slon}. This STT can affect the
dynamics of magnetization $\vec{M}$ of F when its thickness is thin
enough. Theoretical studies\cite{Slon,theory,three} show that the
STT $\Gamma$ is proportional to the current with the following form
\begin{equation}\label{torque}
\Gamma \equiv [\frac{d(\vec{M}V)}{dt}]_{STT} =\frac{\gamma\hbar
I}{\mu_0 e } g(P, \vec{m}\cdot \hat{s}) \vec{m}\times (\vec{m}
\times \hat{s})
\end{equation}
where $\vec{m}, \hat{s}$ are the unit vectors of $\vec{M}$ and
the polarization direction (along $\vec{M}_1$) of the current,
respectively. In the expression, $V$, $\hbar$, $\mu_0$, and $e$
denote the volume of F, the Planck constant, the vacuum magnetic
permeability, and the electron charge, respectively. $\gamma=2.21
\times 10 ^5 (rad/s)/(A/m)$ is the gyromagnetic ratio. The exact
microscopic formulation of the STT is still a subject of study and
debate\cite{theory,three}. Different theories differ themselves in
different expressions of function $g$ that depends on the degree of
the polarization $P$ of the current and relative angle between
$\vec{m}$ and $\hat{s}$. All experimental investigations\cite{exp}
so far are consistent with the result of Slonczewski\cite{Slon}
which will be used throughout this study,
\begin{equation}
g(P, \vec{m}\cdot \hat{s})=\frac{4P^{3/2}}{(1+P)^3(3+\vec{m}\cdot
\hat{s})-16P^{3/2}}.
\end{equation}

The magnetization dynamics of a Stoner particle (F in Fig.
1) under an effective magnetic field $\vec{H}_{t}$ and
a polarized current is governed by the so-called modified
Landau-Lifshitz-Gilbert (LLG) equation\cite{theory} with
an additional term (the third term on the right of the
equation below) due to the STT of Eq.~\eqref{torque}
\begin{equation}
\frac{d\vec{M}}{dt}=-\gamma\vec{M} \times \vec{H}_{t}+\alpha
\vec{m} \times \frac{d\vec{M}}{dt} +\gamma a_I\vec{M}\times
(\vec{M} \times \hat{s}),\label{MLLG}
\end{equation}
where $a_I=\hbar Ig/(\mu_0 e M^2V)$ is a dimensionless parameter,
and $\alpha$ is a phenomenological dimensionless damping
constant whose typical value ranges from 0.01 to 0.22 for
Co films\cite{field}. Because the magnitude of $\vec{M}$ does
not change with time for a Stoner particle, $\vec{M}$ can be
described by the polar angle $\theta$ and the azimuthal angle
$\phi$ in the spherical coordinate, and Eq. \eqref{MLLG} can
be rewritten in a dimensionless form
\begin{equation}
(1+\alpha^2)\frac{d\vec{m}}{dt} = - \vec{m} \times \vec{h}_{1}
-\vec{m} \times (\vec{m} \times \vec{h}_{2}), \label{DMLLG}
\end{equation}
where
\begin{align*}
\vec{h}_1 &= \vec{h}_t+\alpha a_I\hat{s},\\
\vec{h}_2 &= \alpha\vec{h}_t- a_I\hat{s}.
\end{align*}
$t$ in Eq. \eqref{DMLLG} is measured in unit of $(\gamma M)^{-1}$.
The magnetization and the magnetic field are in the units of $M$.
the total field $\vec{h}_t=\vec{h}+\vec{h}_i$ includes both
the applied magnetic field $\vec{h}$ and the internal field
$\vec{h}_i$ due to the magnetic anisotropic energy density
$w(\vec{m})$, $\vec{h}_i=-\nabla_{\vec{m}}w(\vec{m})/\mu_0$.
Different particles are characterized by different magnetic
anisotropy. We consider only the uniaxial particles of magnetic
anisotropy $w(\vec{m})=-k m_z^2/2$ with its easy axis along the
z-direction so that and $\vec{h}_i=-\frac{\partial w(\cos\theta)}
{\partial (\cos \theta)}\hat{z}\equiv f(\cos\theta)\hat{z}$.
Let $\hat{e}_r,\ \hat{e}_{\theta},\ \hat{e}_{\phi}$ be the three
spherical unit vectors of $\vec{m}$, so $\vec{h}_i=-k\sin\theta
\cos\theta \hat{e}_{\theta}+k\cos^2\theta\hat{e}_r$. In terms of
$\theta$ and $\phi$, Eq.~\eqref{DMLLG} can be written as
\begin{equation}
\begin{split}
&(1+\alpha^2)\dot{\theta} = h_{t,\phi} +\alpha h_{t,\theta}
+a_I(\alpha s_{\phi}-s_{\theta}),\\
&(1+\alpha^2)\sin\theta\dot{\phi} = \alpha h_{t,\phi} -
h_{t,\theta} - a_I(\alpha s_{\theta}+s_{\phi}).
\end{split}\label{sphe}
\end{equation}
Here $h_{t,\theta}$, $h_{t,\phi}$ and $s_{\theta}$, $s_{\phi}$,
$s_{r}$ are the $\hat{e}_{\theta}$, $\hat{e}_{\phi}$, and $\hat
{e}_r$ components of $\vec{h}_t$ and $\vec{s}$, respectively.

Because $\vec{h}_1$ and $\vec{h}_2$ are in general non-collinear,
the dynamics with the additional STT term in Eq. \eqref{DMLLG}
is very different from that without this term which describes a
Stoner particle only in a magnetic field. The particle energy can
only decrease in a static magnetic field since the field cannot
be an energy source\cite{xrw1}. However, a polarized electric
current can pump energy into the particle through the STT.
Thus, the STT term from even a dc current can be an energy source,
and the dynamical behavior of a Stoner particle under an STT
is much richer than its counterpart in a static magnetic field.
According to Eq. \eqref{DMLLG}, the magnetization undergoes a
precessional motion around field $\vec{h}_1$ and a damping motion
toward a different field $\vec{h}_2$.

The magnetization switching problem for a uniaxial Stoner
particle is as follows: In the absence of both polarized current
and $\vec{h}(=0)$, $\vec{M}$ in F has two stable directions,
$\theta=0$ and $\theta=\pi$, along its easy axis (z-axis).
The goal is to reverse the initial state (say $\theta=0$) to the
target state, $\theta=\pi$, fast by a small polarized electric
current. All studies so far assumed that the current polarization
direction does not vary with time. However, previous
studies\cite{xrw1} on magnetic-field induced magnetization
reversal show that a switching field can be dramatically
reduced if the direction of the field varies properly with time
during a reversal. Thus, it is natural to investigate whether
one can use a current with a proper time-dependent polarization
direction to lower the critical reversal current density.
The precise issue studied here is: For a given system in which
damping constant $\alpha$, anisotropy $f(\cos\theta)$ and
the external magnetic field are fixed, the polarized electric
current can vary its polarizations direction during a reversal
process under the constrain of the constant magnitude
$I$ and constant polarization degree $P$ of the current.
The issue is to minimize the critical current density, and
to find the best shape of polarization direction for the
shortest reversal time.

To simplify the calculations, we consider the case of zero
external magnetic field, $\vec{h}=0$. However, the basic
idea should directly be applicable for non-zero static
external field because $\vec{h}$ can be added to $\vec{h}_i$.
Then Eq.~\eqref{sphe} becomes,
\begin{equation}\begin{split}
&(1+\alpha^2)\dot{\theta} = a_I(\alpha s_{\phi}-s_{\theta})-
\alpha f(\cos\theta) \sin\theta,\\
&(1+\alpha^2)\sin\theta\dot{\phi} = - a_I(\alpha
s_{\theta}+s_{\phi}) +f(\cos\theta)\sin\theta.
\end{split}\label{sphe1}
\end{equation}
$\hat{s}=\hat{s}(t)$ is in general a function of the time.
Different function will lead to different angular
velocities for $\theta$ and $\phi$. Thus, the magnetization
reversal time from the initial state ($\theta=0$) to the
target state ($\theta=\pi$), defined as $T \equiv \int
_0^{\pi} d\theta/\dot{\theta}$, depends on $\hat{s}(t)$.
In order to find the optimal $\hat{s}(t)$ that minimizes
$T$, one only needs $a_I(\alpha s_{\phi}-s_{\theta})$
or $g(P,s_r)(\alpha s_{\phi}-s_{\theta})$ to be maximum
such that $\dot{\theta}$, according to Eq.~\eqref{sphe1},
will be the largest at any $\theta$.
This observation is important and it can be applied to
other function forms of $g$. Because $s_r^2+s_{\theta}^2 +
s_{\phi}^2=1$, the maximum of $g(P, s_r)(\alpha s_{\phi} -
s_{\theta})$ can be obtained from the standard Lagrange
multiplier method in which one introduces $L\equiv g(P,s_r)
(\alpha s_{\phi}-s_{\theta})-\lambda (s_r^2+s_{\theta}^2 +
s_{\phi}^2)$. By setting the partial derivatives of $L$
with respect to $s_i$ (i=r, $\theta$, $\phi$) to zeros,
the maximum of $g(P, s_r)(\alpha s_{\phi}-s_{\theta})$ is
\begin{equation}
[g(P, s_r) (\alpha s_{\phi}-s_{\theta})]_{max}=
\sqrt{1+\alpha^2} G(P),
\end{equation}
where $G(P)=g(P, s_r^*)\sqrt{1-s_r^{*2}}$ and
\begin{align}
&s_r^* = \frac{(1+P)^3}{16P^{3/2}-3(1+P)^3}, \nonumber\\
&s_{\theta}^* =-\frac{1}{\sqrt{1+\alpha^2}}\sqrt{1-s_r^{*2}},
\label{pulse} \\    &s_{\phi}^*
=\frac{\alpha}{\sqrt{1+\alpha^2}}\sqrt{1-s_r^{*2}}.\nonumber
\end{align}
Eq.~\eqref{pulse} gives the optimal polarization direction
which will lead to the shortest switching time under a fixed
current magnitude. Although the optimal $\vec{s}^*$ appears
to depend only on damping constant $\alpha$ and $P$, but not
on $f(\cos \theta)$, it is in fact time-dependent since
$\hat{e}_r,\ \hat{e}_{\theta},\ \hat{e}_{\phi}$ vary with
the time. Furthermore, magnetic anisotropy
$f(\cos \theta)$ shall influence the evolution of $\vec{m}$
which in turn influences the time-dependence of $\vec{s}^*$.
Thus, if they were to change $f(\cos \theta)$ and nothing
else, the time-dependent $\vec{s}^*$ would be different.

Under the optimal design of Eq.~\eqref{pulse}, $\theta(t)$ and
$\phi(t)$ will satisfy, respectively,
\begin{equation}\label{dottheta} \dot{\theta} =
\frac{\hbar I }{\mu_0 e M^2 V} \frac{G(P)}{\sqrt{1+\alpha^2}}
-\alpha f(\cos\theta)\sin\theta/(1+\alpha^2),
\end{equation}
and
\begin{equation}\label{dotphi}
\dot{\phi} =f(\cos\theta)/(1+\alpha^2).
\end{equation}
For $w(\vec{m})=-km_z^2/2$, it is straightforward to integrate
Eq.~\eqref{dottheta}, and obtain the reversal time $T$ from
$\theta=0$ to $\theta=\pi$,
\begin{equation}\label{time}
T=\frac{2}{k} \frac{(\alpha^2+1)\pi}{ \sqrt{
4(\alpha^2+1)\hbar^2G^2(P)I^2/(\mu_0 eM^2Vk)^2-\alpha^2}}.
\end{equation}
In the weak damping limit ($\alpha\rightarrow 0$) or large current
limit ($I \rightarrow \infty$), $T \propto \pi/I$.

For a uniaxial model, the critical switching current or current
density $J_c$ can be obtained by setting $\dot{\theta}=0$ in
Eq.~\eqref{dottheta}. This is because $\dot{\theta}$ cannot be
negative if the magnetization of a uniaxial particle moves from
$\theta=0$ to $\theta=\pi$. Therefore, the first term in
Eq.~\eqref{dottheta} must exceed the second term due to magnetic
anisotropy for all ${\theta}$'s ($\in [0,\pi]$) in a reversal.
This simplicity for a uniaxial model comes from
$\phi$-independence of Eq.~\eqref{dottheta}.
Since Eq.~\eqref{dottheta} is the largest velocity for an
arbitrary ${\theta}$ under the best choice of the polarization
direction of a current, the critical current should be the one
when the smallest $\dot{\theta}$ (for all $\theta$) is zero.
The critical reversal current density in our case is
\begin{equation}\label{newc}
J_c =\frac {\mu_0 e M^2 d}{\hbar
G(P)}\frac{\alpha}{\sqrt{1+\alpha^2}}Q.
\end{equation}
Here the current density is defined as $J=I/A$ with $A$ being
the cross-section area and $d$ being the thickness of F.
$Q\equiv max\{f(\cos\theta)\sin\theta\}$ for $\theta\in [0,\pi]$.
$Q=k/2$ for $w(\vec{m})=-km_z^2/2$. One notices, from the
derivation of Eq.~\eqref{newc}, that the assumptions of
time-independence of $I$ and $P$ are not essential for $J_c$
as long as STT is proportional to $I$ and $g$ in Eq. (1).

To see how much the critical switching current density is reduced,
let us compare the critical current density of Eq.~\eqref{newc}
with those in the previous schemes where the current
polarization direction is fixed. When the polarization direction
$\hat{s}$ is parallel to the easy axis of the uniaxial magnet F
(the critical current density in the perpendicular case is larger),
the critical current density for the same anisotropy as that
of Eq.~\eqref{newc} is\cite{Slon,sun,lee}
\begin{equation}\label{oldc}
J_c =\frac {\mu_0 e M^2 d}{\hbar g(P, 1)}\alpha k.
\end{equation}
Fig. 2 is the plot of $J_c$ versus damping constant $\alpha$ for
$P=0.4$. The dashed line is that for Eq.~\eqref{oldc}, and the
solid line is the result of our new strategy which saturates to
a constant at large $\alpha$ limit. At $\alpha=0.1$, $J_c$ in the
new strategy is about one fourth of that given by Eq. \eqref{oldc}.
The difference between Eqs. \eqref{newc} and \eqref{oldc} depends
on the degree of polarization, $P$. Fig.3 is $J_c$ vs. $P$ at
$\alpha=0.1$. It should be pointed out that zero $J_c$ in Eq.
\eqref{newc} at $P=1$ is an artificial result originated from the
divergence of $g(1,x)$ at $x=-1$ in the Slonczewski's theory\cite{Slon}.
This divergence is removed in other formulations of $g$\cite{three}.
\begin{figure}[htbp]
 \begin{center}
\scalebox{0.75}[0.75]{\includegraphics[angle=0]{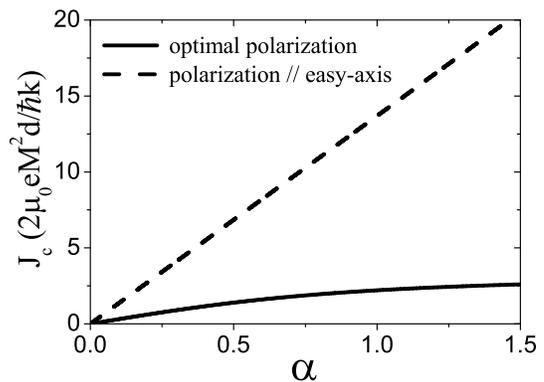}}
 \end{center}
\caption{\label{fig2} $J_c$
vs. $\alpha$ for $P=0.4$ and uniaxial model
of $w(\vec{m})=-k m_z^2/2$.}
\end{figure}

\begin{figure}[htbp]
 \begin{center}
\scalebox{0.75}[0.75]{\includegraphics[angle=0]{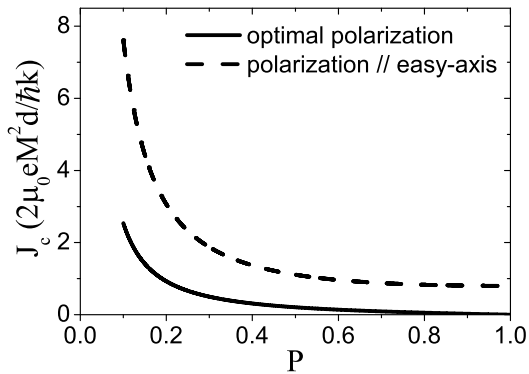}}
 \end{center}
\caption{\label{fig3} $J_c$ vs. $P$ at $\alpha=0.1$. The rest of
system parameters are the same as that of Fig. 2.}
\end{figure}

Although our results are obtained for uniaxial Stoner particles
with a specific function of $g$, the basic ideas and approaches
can in principle be generalized to the non-uniaxial cases with
other $g$-functions. Also, almost all experiments so far used a
set-up like that illustrated in Fig.1. However, the formulation
of an STT induced magnetization reversal does not really rely
on the set-up. In general, the same formulation could be
applied to a Stoner particle through which a spin current pass.
Therefore, advance in generating the pure spin current
(with charge current being zero), one of current topics in
so-called spin Hall effect, could lead to a new set of
experimental set-up beyond the prototype illustrated
in Fig. 1 in the STT experiments and the STT applications.
In the prototype of Fig. 1, it happens that the current of
spin-polarized electrons is generated by one of the magnets.

The realization of the results reported here depends largely
on our ability in generating a designed spin-polarized electric
current. In some sense, the present work converts the issue
of critical current to the issue of generating an arbitrary
polarized electric current. Any breakthrough in the front of
spin current generation shall lead to the great leap forward
in magnetization minipulation. One possible way to generate a
desired polarized electric current is through controlling
magnetization $\vec{M}_1$ of F1 in Fig. 1 by other means.
As it was explained early, the polarization direction of the
electric current is parallel to $\vec{M}_1$ in an experimental
set-up illustrated in Fig. 1. Thus, time-dependence of the
polarization of the current is the same as that of $\vec{M}_1$.
However, since the response time of electrons to an electric
current is usually much faster than that of the magnetization,
it may be an experimental challenge to create a required
time-dependent spin polarization by the method, especially
when the change of polarization is very fast.

In conclusion, we have showed that a proper time-dependent
polarized electric current can dramatically reduce the
critical current density needed to reverse a magnetization.
An optimal time-dependent current polarization is obtained
such that the magnetization reversal time is the shortest
for uniaxial Stoner particles.

{\it{Acknowledgments}--}This work is supported by UGC, Hong Kong,
through RGC CERG grants (\#603106).

\end{document}